\documentclass[english,aps,preprint]{revtex4}
\usepackage[T1]{fontenc}
\usepackage[latin9]{inputenc}
\usepackage{textcomp}
\usepackage{graphicx}

\usepackage{babel}
\begin{document}

\title{Affect of film thickness on the blue photoluminescence from ZnO}

\author{J.C. Moore}
\email{moorejc@coastal.edu}

\author{L.R. Covington}
\author{R. Stansell}

\affiliation{Department of Chemistry and Physics, Coastal Carolina University,
Conway, SC 29528}
\begin{abstract}
Zinc oxide (ZnO) films having various thicknesses were synthesized on sapphire substrates by thermal oxidation of Zn-metallic films in air ambient. X-ray diffraction (XRD) spectra indicate that the resulting films possess a polycrystalline hexagonal wurtzite structure without preferred orientation. For films having a thickness of 200 nm, crystal grain size was observed to decrease with increasing annealing temperature up to 600\textdegree{}C, and then increase at higher temperatures. Thicker films demonstrated a modest increase in grain size as temperature increased from 300\textdegree{}C to 1200\textdegree{}C. The influence of film thickness on the optical properties was investigated using room temperature photoluminescence (PL). Specifically, PL spectra indicate four emission bands: excitonic ultraviolet, blue, and deep-level green and yellow emission. The strongest UV emission and narrowest full width at half maximum (0.09 eV) was observed for films having a thickness of 200 nm and annealed at low temperature (300 \textdegree{}C). The ratio of deep-level green emission to UV excitonic emission was observed to decrease with decreasing annealing temperature, which is attributed to the generation of fewer oxygen vacancies (V$_{o}^{+}$) and interstitial oxygen ions (O$_{i}^{-}$) in the bulk. As film thickness decreased, we observed the emergence of blue emission and a significant red shift (0.15 eV) in the bandgap. The emergence of blue emission and the corresponding decrease in emission associated with bulk defects when depletion width grows relative to the bulk suggests that the origin of the blue emission is related to the negatively charged Zinc interstitials (V$_{Zn}^{-}$) found within the deletion region near the interface.
\end{abstract}
\maketitle

\section{Introduction}
Zinc oxide (ZnO) is a wide bandgap semiconductor that has attracted
a great deal of attention with demonstrated applications in ultraviolet
(UV) light detection, air-quality monitoring, missile warning systems,
gas detection, and utilization as light-emitting diodes. Thin films
of ZnO have been grown via a variety of methods, including the sol-gel
method, molecular beam epitaxy, chemical vapor deposition, and the
thermal oxidation of Zn-metal films.\cite{Ozgur2005} Specifically
with regards to thermal oxidation, Cho \emph{et al.} demonstrate growth
of un-doped ZnO films having a polycrystalline wurtzite structure,
where grain size is seen to increase with annealing temperature.\cite{Cho1999}
The photoluminescence (PL) spectrum for ZnO typically includes a broad
ultraviolet (UV) band emission attributed to excitonic binding energies
and green and yellow band emission, due to various defect states.\cite{Wu2001,Wanga2006,Fu1998}
In several instances, blue-band emission has also been reported.\cite{Zhang2002,Zhao2008,Zhao2004,Fujihara2004}

Wu \emph{et al.} argue that deep-level green and yellow emission corresponds
to a recombination of a delocalized electron close to the conduction
band with a deeply trapped hole in the single ionized oxygen vacancy (V$_{o}^{+}$)
and the single negatively charged interstitial oxygen (O$_{i}^{-}$) ion centers,
respectively.\cite{Wu2001} There has been much debate on the cause
of defects leading to specific PL emission, and several possibilities
have been suggested such as growth environment, annealing temperature,
and film thickness. Wang \emph{et al.} report weak deep level PL emission
bands compared to UV emission, and green band intensity that increases
at higher temperatures when Zn-metal films are oxidized in air.\cite{Wang2003}
However, Chen \emph{et al.} show decreasing deep level band intensities
with increasing temperature when films are oxidized in oxygen ambient
via a two-step process.\cite{Chen2002} These seemingly conflicting
reports suggest that the oxidizing agent and final annealing temperature
is critical in determining PL behavior.

The origin of blue emission is controversial. Lee \emph{et al.} describe
strong blue emission possibly caused by stress resulting from the
volume expansion of the ZnO transformed from Zn during high treatment
temperature.\cite{Lee2008} Whereas Zhang \emph{et al.} have shown
that the intensity of the blue PL peak is strongly dependent on the
oxygen pressure.\cite{Zhang2002} They conclude that one source of
blue level emission is from the electron transition from the shallow
donor of oxygen vacancies to the valence band, while another is electron
transmission from the shallow donor level of zinc interstitials to
the valence band. However, Wu \emph{et al.} argue that blue emission
is related to Zn interstitials found within the depletion region.
They argue that a large depletion region relative to the bulk results in
blue emission, since bulk-related defects associated with deep-level
emission would dominate except when bulk volume is comparable to that
of the depletion region.\cite{Wu2001}

We investigate the effects of film thickness on PL emission for ZnO grown via thermal oxidation of Zn-metal films. Specifically, if green and blue emission result from defect-related energy levels in the bulk and depletion region, respectively, as suggested by Wu \emph{et al.},\cite{Wu2001} and
we decrease the bulk to depletion region ratio via a decrease in thickness, then we should observe a corresponding decrease in the green to blue emission ratio.

\section{Experiments}
In this study, Zinc films were grown on\emph{ c}-plane sapphire substrates using direct current sputter deposition without reactive gas. The sputter cathode used was a 1" diameter 99.99\% purity zinc target mounted on a water-cooled stage. A turbomolecular pump maintained a background pressure of \textasciitilde{}$1\times10^{-6}$ mbars before deposition. During deposition, an argon pressure of \textasciitilde{}$2\times10^{-2}$
mbars was maintained via a metal leak valve and pump throttling. Deposition times ranged from 10-60 minutes at sputter power between 15-30 W. All resulting Zn films were initially oxidized by thermal annealing in an open air muffle furnace at a temperature of 300\textdegree{}C for over 24 hours. Some films where then re-annealed at temperatures of 600\textdegree{}C, 900\textdegree{}C, and 1200\textdegree{}C for two hours. After thermal oxidation, all films where removed from the furnace and allowed to cool in air ambient.

Structural properties of the resulting films where investigated using x-ray diffraction (XRD). XRD spectra indicate that after annealing, the resulting ZnO films possess a polycrystalline hexagonal wurtzite structure without preferred orientation. Grain size was calculated from the Scherrer formula and XRD spectra, and confirmed via atomic force microscopy (AFM, Anfatec Level).\cite{Cullity1978}  Surface morphology and film thickness was measured using AFM. Film thickness was confirmed via reflectometry using a broad spectrum fluorescent source and a UV-Vis spectrophotometer. Post-annealed ZnO films demonstrated thicknesses ranging from \textasciitilde{}200 nm to \textasciitilde{}600 nm. Bandgap energies were determined by absorption band edge using the same spectrophotometer with an integrated tungsten-deuterium source. Photoluminescence spectra where acquired at room temperature using a HeCd laser (325 nm) as an excitation source at a power P = 0.3 W/cm$^{2}$. 

\begin{figure}%
\includegraphics[scale=0.8]{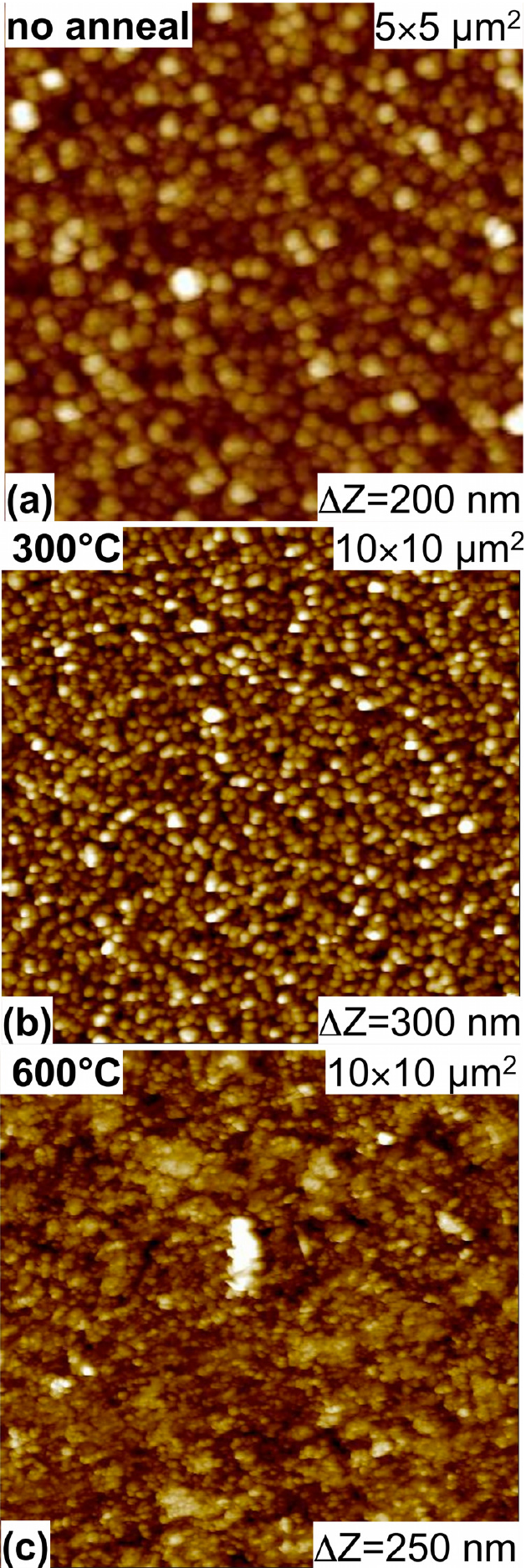}
\caption{AFM topography images of (a) Zn-metal films before oxidation (grayscale range = 200 nm) and the resulting polycrystalline ZnO films annealed at (b) 300\textdegree{}C (grayscale range = 300 nm), and (c) 600\textdegree{}C (grayscale range = 300 nm). Film thickness is 200 nm. Crystal grain size is observed to decrease with increasing annealing temperature.\label{fig:afm}}
\end{figure}

\section{Results and discussion}
The evolution of the surface morpholgy with increasing annealing temperature for 200 nm thick ZnO films is shown in the AFM images presented in fig. \ref{fig:afm}(a-c) The as-grown zinc film demonstrates a high surface roughness and approximately 100 nm diameter protrusions [fig. \ref{fig:afm}(a)]. Figures \ref{fig:afm}(b-c) show the  surface morphology of resulting ZnO films annealed at 300\textdegree{}C and 600\textdegree{}C. There is very little change in the underlying characteristics between zinc-metallic films and ZnO films annealed at 300\textdegree{}C; however, surface roughness is observed to increase and protrusions grow in size by approximately 50 nm. This is consistent with previous studies, where Gupta \emph{et al.} show that the preferred orientation of ZnO thermally oxidized on glass can depend on the Zn film texture and oxidizing agent.\cite{Gupta2002} As shown in fig. \ref{fig:afm}(c), a decrease in protrusion diameter and surface roughness is observed with increasing temperature. There is little change in surface morphology observed at higher temperatures (not shown). Specifically, protrusion size does not significantly change.

Grain size was characterized by both AFM and XRD. For films having thicknesses of 400 nm and 600 nm, grain size was observed to increase with increasing annealing temperature from 300\textdegree{}C to 1200\textdegree{}C. Interestingly, for 200 nm thick films, grain size decreased with increasing temperature up to a certain point. Figures \ref{fig:afm}(b-c) show a decrease in protrusion diameter from approximately 150 nm to approximately 100 nm at annealing temperatures of 300\textdegree{}C and 600\textdegree{}C, respectively, with no further significant change in size as temperature was further increased. Grain size, as determined by the full width at half maximum in the XRD spectra (not shown), was also found to decrease with increasing temperature up to 600\textdegree{}C, with relatively small increases at higher temperatures, which is consistent with AFM measurements. This observation appears inconsistent with some reports in the literature for thermally oxidized ZnO films.\cite{Cho1999, Wang2003} This discrepancy may be the result of our two-step thermal annealing process, where metallic zinc films are all initially oxidized at low temperature, the differences in studied temperature regimes, and the variations in film thickness. Furthermore, our metallic zinc films display a significantly different texture and larger initial particle size in comparison to films grown via other methods.

\begin{figure}%
\includegraphics[scale=0.8]{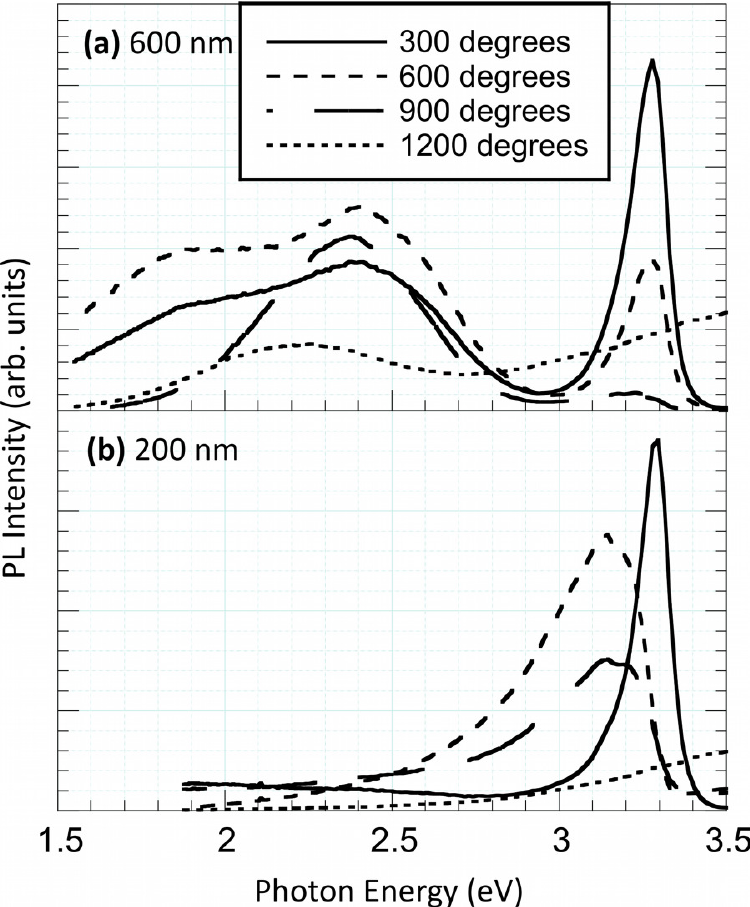}
\caption{PL spectra of ZnO films grown at 300\textdegree{}C, 600\textdegree{}C,
900\textdegree{}C and 1200\textdegree{}C at thicknesses of (a) 600
nm and (b) 200 nm. For thicker films, increasing green band emission
relative to UV emission is seen with increasing temperature. Thinner
films demonstrate little green band emission; however, a significant
redshift in UV emission and asymmetrical band broadening is observed
at higher temperatures.\label{fig:PL-spectra}}
\end{figure}

Figure \ref{fig:PL-spectra} shows the PL spectra of ZnO films with thickness of (a) 600 nm and (b) 200 nm thermally annealed at various
temperatures. These spectra all show asymmetric broad bands in the yellow-green region. A more narrow band in the UV is observed in most cases, which is associated with excitonic emission; however, the 200 nm film demonstrates an asymmetric and broad band in the blue-UV region at higher temperatures. For all thicknesses, the strongest UV excitonic emission is observed at low temperature, with decreasing UV emission observed with increasing annealing temperature. Interestingly, for thinner films, increasing temperature results in a significant redshift (0.15 eV) in the UV excitonic peak and an asymmetrical peak broadening. 

\begin{figure}%
\includegraphics[scale=0.8]{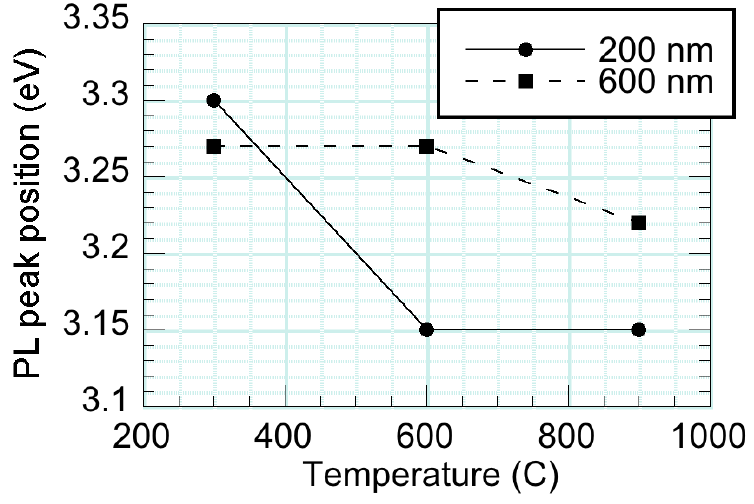}
\caption{Plot of PL peak position with temperature for films having a thickness of 200 nm (solid line) and 600 nm (dashed line). A red shift in UV peak position is observed with increasing temperature for both thicknesses.\label{fig:uv}}
\end{figure}

Figure \ref{fig:uv} shows a plot of the UV peak position with temperature. Very little change in UV peak position is observed for films having a thickness of 600 nm. However, a significant 0.15 eV red shift in UV peak position is observed from 300\textdegree{}C to 600\textdegree{}C, with a negligible shift observed from 600\textdegree{}C to 900\textdegree{}C. This change in UV peak position could be explained by a corresponding shift in bandgap energy, which would be consistent with transmission studies of ZnO films having varying thickness and grown via the sol-gel method.\cite{Jain2007} 

\begin{figure}%
\includegraphics[scale=0.8]{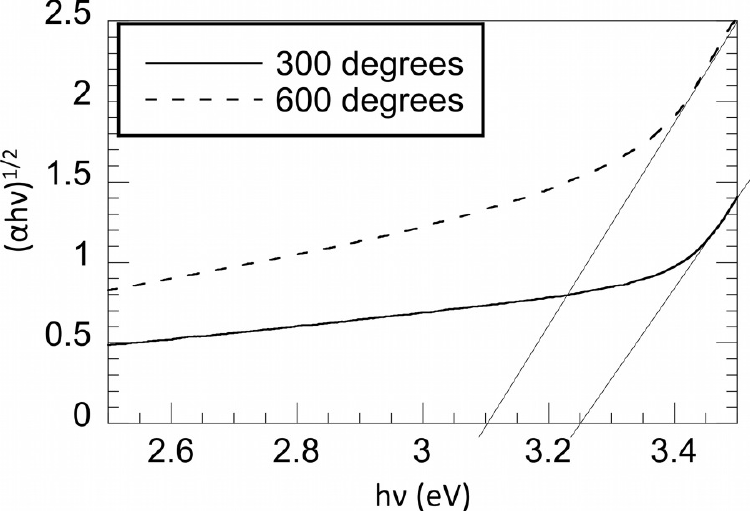}
\caption{Tauc plot obtained from absorption measurements for 200 nm thick ZnO
films annealed at 300\textdegree{}C and 600\textdegree{}C. Indirect
bandgap energies are obtained from the \emph{y}-intercept of the extrapolation
for the linearly increasing region. Bandgap energies are observed
to redshift approximately 0.15 eV with increasing temperature, which
is consistent with observed shifts in UV excitonic emission. \label{fig:Tauc-plot}}
\end{figure}

To determine whether changes in the bandgap were responsible for the observed red shift in UV emission, we measured bandgap energies. Figure
\ref{fig:Tauc-plot} shows the Tauc plot obtained from the absorption spectra for thin films annealed at 300\textdegree{}C and 600\textdegree{}C.\cite{Tauc1968} Absorption measurements indicate an indirect bandgap that redshifts from 3.25 eV to 3.10 eV with increasing annealing temperature. No further redshift in bandgap was observed with increasing temperature past 600\textdegree{}C (not shown). This shift in bandgap energy is consistent with the observed shift in UV emission peak in the PL spectra shown in fig. \ref{fig:PL-spectra}(b), and consistent with bandgap shifts with decreasing film thickness reported in the literature.\cite{Jain2007} Wu \emph{et al.} and Dijken \emph{et al.} both demonstrate that an increase in particle size should result in a redshift in energies, which appears inconsistent with our results.\cite{Wu2001, Dijken2000} However, these studies discuss systems where quantum size effects become relevant, and the particle sizes in this study are sufficiently large that the shift in bandgap can not be explained via a similar mechanism. Jain \emph{et al.} speculate that this red shift is the result of an increase in interstitial zinc atoms.\cite{Jain2007}

\begin{figure}
\includegraphics[scale=0.8]{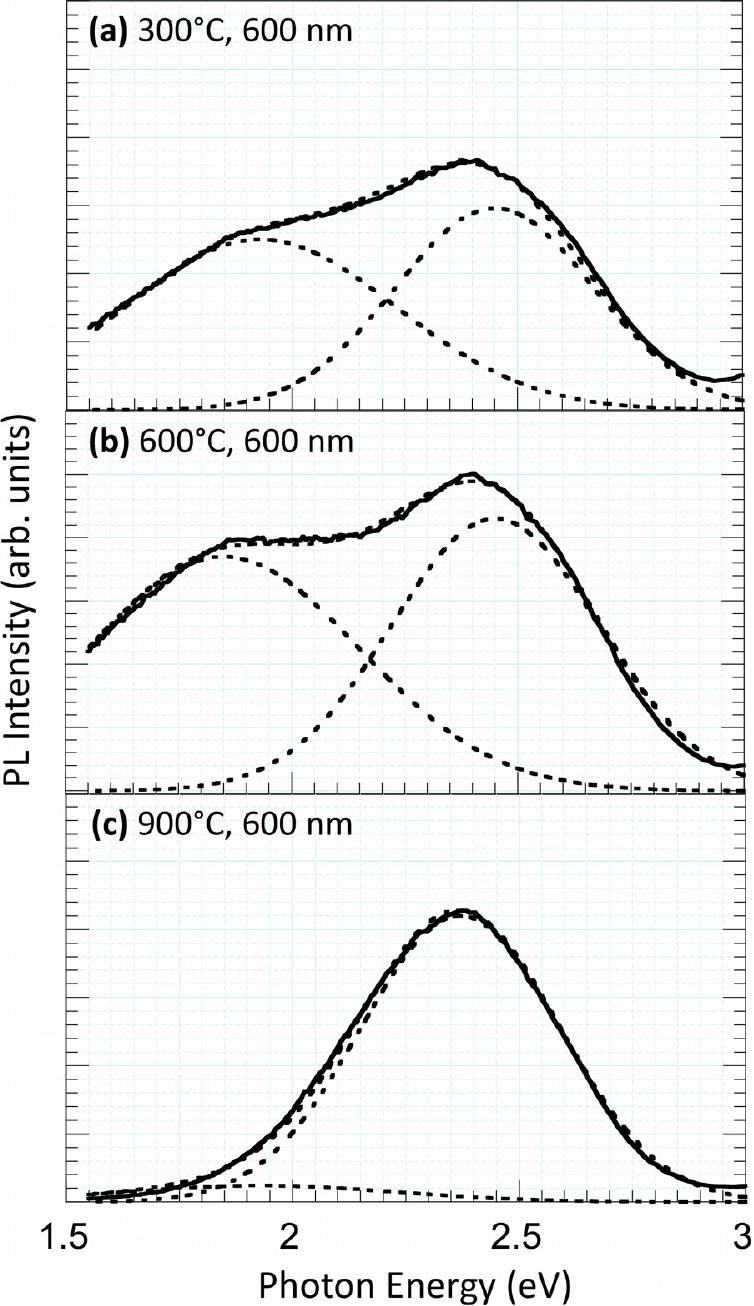}
\caption{PL spectra of ZnO films grown to a thickness of 600 nm and thermally annealed at (a) 300\textdegree{}C, (b) 600 \textdegree{}C, and (c) 900 \textdegree{}C. Solid lines indicate measured PL emission. Dashed lines correspond to fits derived from linear combination of Gaussian peaks centered on yellow and green energies.\label{fig:green}}
\end{figure}

As shown in fig. \ref{fig:PL-spectra}, 600 nm films demonstrate increasing yellow-green band intensity with increasing annealing temperature when compared to UV emission. This is consistent with previous research showing higher temperature results in a rapid increase in V$_{o}^{+}$ and O$_{i}^{-}$ ion centers.\cite{Wang2003, Wu2001} The broad and asymmetrical yellow-green peak observed in fig. \ref{fig:PL-spectra}(a) for the 600 nm film can be Gaussian divided into two bands in the yellow and green region. Figure \ref{fig:green}(a-c) shows the PL spectra (solid lines) for films annealed at various temperatures with a model consisting of the linear combination of two Gaussian peaks centered on yellow and green energies (dashed lines). We observe a small red shift (0.5 eV) in green band emission without a corresponding shift in yellow emission, which Wang \emph{et al.} attribute to differences in the annealing temperature contribution to the respective defects responsible.\cite{Wang2003} Also, the green to yellow ratio increases dramatically at annealing temperatures above 600\textdegree{}C. 

In contrast, 200 nm thick films demonstrate little green and yellow band emission at any temperature [see fig. \ref{fig:PL-spectra}(b)]. Deep level emission is attributed to bulk defects, therefore it is possible that decreased bulk volume results in the formation of relatively fewer deep-level states. If green and yellow emission results from the recombination of a delocalized electron close to the conduction band with a deeply trapped hole in the V$_{o}^{+}$ and O$_{i}^{-}$ centers in the bulk, respectively, then a decrease in film thickness would decrease the bulk with respect to the depletion region, resulting in weaker bulk-related, deep-level emission.\cite{Wu2001} Reaction kinetics could also contribute, with thinner films having shorter diffusion paths for reactive oxygen species during oxidation.\cite{Aida2006} Thin films would therefore demonstrate fewer  V$_{o}^{+}$ and O$_{i}^{-}$ ions at any annealing temperature, as observed.

\begin{figure}
\includegraphics[scale=0.8]{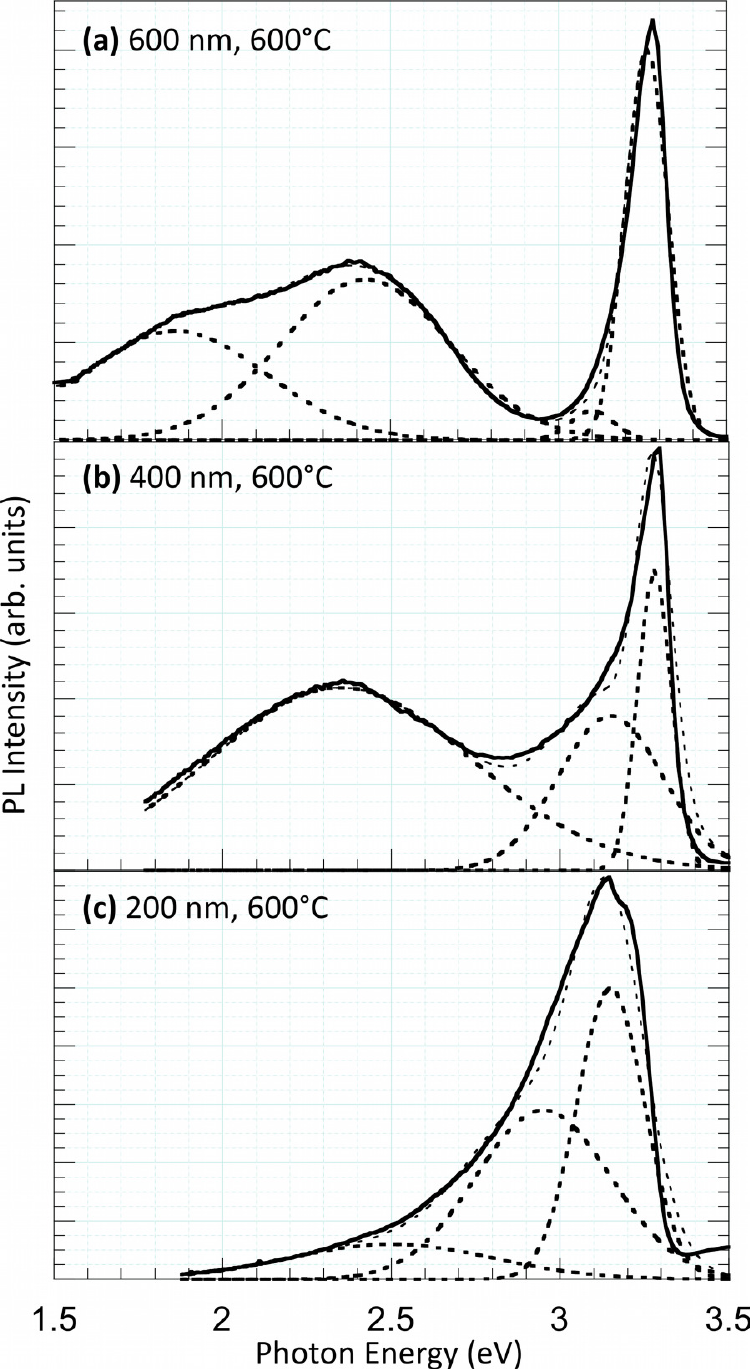}
\caption{PL spectra of ZnO films annealed at 600\textdegree{}C and grown to
thicknesses of approximately (a) 600 nm, (b) 400 nm, and (c) 200 nm.
Solid lines indicate measured PL emission. Dashed lines correspond
to fits derived from linear combination of Gaussian peaks centered
on yellow, green, blue, and UV energies. The presence of blue emission
could explain the observed asymmetric shape about the UV peak. Furthermore,
the ratio of green to blue emission decreases with decreasing film
thickness.\label{fig:Gauss-fits}}
\end{figure}

Figure \ref{fig:Gauss-fits} shows PL spectra (solid lines) for ZnO films annealed at 600\textdegree{}C having three thicknesses: approximately
(a) 600 nm, (b) 400 nm, and (c) 200 nm. To determine the affect of film thickness on the blue emission, we modeled overall PL emission
as the linear combination of four Gaussian peaks centered on yellow, green, blue, and UV energies (dashed lines). The UV excitonic emission is observed as expected considering the bandgap energies of the films. Specifically, the UV peak for the 200 nm film {[}fig. \ref{fig:Gauss-fits}(c){]} is redshifted in a manner consistent with the observed shift in bandgap (see fig. \ref{fig:Tauc-plot}).

The presence of blue emission explains the observed asymmetric shape of the dominant blue-UV peak for all thicknesses. As shown in fig. \ref{fig:Gauss-fits}(a-c), the ratio of green-to-blue emission decreases with decreasing film thickness. The PL spectra for the 200 nm thick film [fig. \ref{fig:Gauss-fits}(c)] exhibited the most dramatic blue band emission and very low green band emission. Similarly, Wu \emph{et al.} found that the intensities of the green and yellow cathodoluminescence peaks were strongly affected by the width of the free-carrier depletion region near the surface.\cite{Wang2003} They argue that single ionized oxygen vacancies exist only in the bulk, so the magnitude of the depletion region in relation to the bulk directly affects the intensity of the green emission in the cathodoluminescence spectrum. It has been suggested that blue emission results from zinc interstitials found in the depletion region, so an approximately 400 nm blue emission should only be observed in a sample with a wide depletion region relative to the bulk. Otherwise, deep-level green emission will dominate.\cite{Zhao2004,Zhao2008}  Therefore, it is not surprising that we would observe the emergence of blue emission with decreasing film thickness, since limiting thickness lowers the ratio of bulk to depletion region.

Temperature and grain size may also contribute, since no blue emission is observed for 200 nm films annealed at 300\textdegree{}C [see fig. \ref{fig:PL-spectra}(b)]. As discussed, we observe larger grain size via XRD and AFM for 200 nm films annealing at 300\textdegree{}C, resulting in a larger bulk to depletion region ratio, which could contribute to weaker blue emission. As with the green emission, reaction kinetics associated with temperature may also contribute, were thinner films would have shorter diffusion paths for reactive oxygen species during oxidation, and lower temperatures would slow the Zn/O$_{2}$ reaction. The result could be fewer zinc interstitials, which would manifest as weaker blue emission and a red shift in both blue and UV emission, as is observed.

\section{Conclusion}
In summary, we have investigated the PL emission from ZnO grown on sapphire substrates via the thermal oxidation of Zn-metal films at
various temperatures and thicknesses. PL spectra indicate four emission bands, excitonic ultraviolet (UV), blue, and deep-level green and
yellow emission. Films annealed at lower temperature exhibited the strongest UV emission intensity and lowest blue and deep-level emission.
An increasing ratio of deep-level green to excitonic UV emission was observed with increasing annealing temperature, which may be attributed to the generation of oxygen vacancies and interstitial oxygen ions in the bulk. The ratio of green to blue band emission decreased with decreasing film thickness, suggesting that the origin of the blue emission is related to Zinc interstitials found within the deletion region near the surface.

\begin{acknowledgments}
The authors would like to thank Everett Carpenter and Mikhail Reshchikov from Virginia Commonwealth University for providing access to their
XRD and PL systems, respectively. This work was funded by NSF DMR \#1104600.
\end{acknowledgments}


\begin{thebibliography}{18}%
\makeatletter
\providecommand \@ifxundefined [1]{%
 \@ifx{#1\undefined}
}%
\providecommand \@ifnum [1]{%
 \ifnum #1\expandafter \@firstoftwo
 \else \expandafter \@secondoftwo
 \fi
}%
\providecommand \@ifx [1]{%
 \ifx #1\expandafter \@firstoftwo
 \else \expandafter \@secondoftwo
 \fi
}%
\providecommand \natexlab [1]{#1}%
\providecommand \enquote  [1]{``#1''}%
\providecommand \bibnamefont  [1]{#1}%
\providecommand \bibfnamefont [1]{#1}%
\providecommand \citenamefont [1]{#1}%
\providecommand \href@noop [0]{\@secondoftwo}%
\providecommand \href [0]{\begingroup \@sanitize@url \@href}%
\providecommand \@href[1]{\@@startlink{#1}\@@href}%
\providecommand \@@href[1]{\endgroup#1\@@endlink}%
\providecommand \@sanitize@url [0]{\catcode `\\12\catcode `\$12\catcode
  `\&12\catcode `\#12\catcode `\^12\catcode `\_12\catcode `\%12\relax}%
\providecommand \@@startlink[1]{}%
\providecommand \@@endlink[0]{}%
\providecommand \url  [0]{\begingroup\@sanitize@url \@url }%
\providecommand \@url [1]{\endgroup\@href {#1}{\urlprefix }}%
\providecommand \urlprefix  [0]{URL }%
\providecommand \Eprint [0]{\href }%
\providecommand \doibase [0]{http://dx.doi.org/}%
\providecommand \selectlanguage [0]{\@gobble}%
\providecommand \bibinfo  [0]{\@secondoftwo}%
\providecommand \bibfield  [0]{\@secondoftwo}%
\providecommand \translation [1]{[#1]}%
\providecommand \BibitemOpen [0]{}%
\providecommand \bibitemStop [0]{}%
\providecommand \bibitemNoStop [0]{.\EOS\space}%
\providecommand \EOS [0]{\spacefactor3000\relax}%
\providecommand \BibitemShut  [1]{\csname bibitem#1\endcsname}%
\let\auto@bib@innerbib\@empty
%</preamble>
\bibitem [{\citenamefont {{\"O}zg{\"u}r}\ \emph {et~al.}(2005)\citenamefont
  {{\"O}zg{\"u}r}, \citenamefont {Alivov}, \citenamefont {Liu}, \citenamefont
  {Teke}, \citenamefont {Reshchikov}, \citenamefont {Do{\u g}an}, \citenamefont
  {Avrutin}, \citenamefont {Cho},\ and\ \citenamefont {Morko{\c
  c}}}]{Ozgur2005}%
  \BibitemOpen
  \bibfield  {author} {\bibinfo {author} {\bibfnamefont {U.}~\bibnamefont
  {{\"O}zg{\"u}r}}, \bibinfo {author} {\bibfnamefont {Y.~I.}\ \bibnamefont
  {Alivov}}, \bibinfo {author} {\bibfnamefont {C.}~\bibnamefont {Liu}},
  \bibinfo {author} {\bibfnamefont {A.}~\bibnamefont {Teke}}, \bibinfo {author}
  {\bibfnamefont {M.~A.}\ \bibnamefont {Reshchikov}}, \bibinfo {author}
  {\bibfnamefont {S.}~\bibnamefont {Do{\u g}an}}, \bibinfo {author}
  {\bibfnamefont {V.}~\bibnamefont {Avrutin}}, \bibinfo {author} {\bibfnamefont
  {S.-J.}\ \bibnamefont {Cho}}, \ and\ \bibinfo {author} {\bibfnamefont
  {H.}~\bibnamefont {Morko{\c c}}},\ }\href@noop {} {\bibfield  {journal}
  {\bibinfo  {journal} {J. Appl. Phys.}\ }\textbf {\bibinfo {volume} {98}},\
  \bibinfo {pages} {1} (\bibinfo {year} {2005})}\BibitemShut {NoStop}%
\bibitem [{\citenamefont {Cho}\ \emph {et~al.}(1999)\citenamefont {Cho},
  \citenamefont {Ma}, \citenamefont {Kim}, \citenamefont {Sun}, \citenamefont
  {Wong},\ and\ \citenamefont {Ketterson}}]{Cho1999}%
  \BibitemOpen
  \bibfield  {author} {\bibinfo {author} {\bibfnamefont {S.}~\bibnamefont
  {Cho}}, \bibinfo {author} {\bibfnamefont {J.}~\bibnamefont {Ma}}, \bibinfo
  {author} {\bibfnamefont {Y.}~\bibnamefont {Kim}}, \bibinfo {author}
  {\bibfnamefont {Y.}~\bibnamefont {Sun}}, \bibinfo {author} {\bibfnamefont
  {G.}~\bibnamefont {Wong}}, \ and\ \bibinfo {author} {\bibfnamefont
  {J.}~\bibnamefont {Ketterson}},\ }\href@noop {} {\bibfield  {journal}
  {\bibinfo  {journal} {Appl. Phys. Lett.}\ }\textbf {\bibinfo {volume} {75}},\
  \bibinfo {pages} {2761} (\bibinfo {year} {1999})}\BibitemShut {NoStop}%
\bibitem [{\citenamefont {Wu}\ \emph {et~al.}(2001)\citenamefont {Wu},
  \citenamefont {Siu}, \citenamefont {Fu},\ and\ \citenamefont {Ong}}]{Wu2001}%
  \BibitemOpen
  \bibfield  {author} {\bibinfo {author} {\bibfnamefont {X.}~\bibnamefont
  {Wu}}, \bibinfo {author} {\bibfnamefont {G.}~\bibnamefont {Siu}}, \bibinfo
  {author} {\bibfnamefont {C.}~\bibnamefont {Fu}}, \ and\ \bibinfo {author}
  {\bibfnamefont {H.}~\bibnamefont {Ong}},\ }\href@noop {} {\bibfield
  {journal} {\bibinfo  {journal} {Appl. Phys. Lett.}\ }\textbf {\bibinfo
  {volume} {78}} (\bibinfo {year} {2001})}\BibitemShut {NoStop}%
\bibitem [{\citenamefont {Wanga}\ \emph {et~al.}(2006)\citenamefont {Wanga},
  \citenamefont {Zu}, \citenamefont {Zhu},\ and\ \citenamefont
  {Wang}}]{Wanga2006}%
  \BibitemOpen
  \bibfield  {author} {\bibinfo {author} {\bibfnamefont {Z.}~\bibnamefont
  {Wanga}}, \bibinfo {author} {\bibfnamefont {X.}~\bibnamefont {Zu}}, \bibinfo
  {author} {\bibfnamefont {S.}~\bibnamefont {Zhu}}, \ and\ \bibinfo {author}
  {\bibfnamefont {L.}~\bibnamefont {Wang}},\ }\href@noop {} {\bibfield
  {journal} {\bibinfo  {journal} {Physica E}\ }\textbf {\bibinfo {volume}
  {35}},\ \bibinfo {pages} {199} (\bibinfo {year} {2006})}\BibitemShut
  {NoStop}%
\bibitem [{\citenamefont {Fu}\ \emph {et~al.}(1998)\citenamefont {Fu},
  \citenamefont {Guo}, \citenamefont {Lin},\ and\ \citenamefont
  {Liao}}]{Fu1998}%
  \BibitemOpen
  \bibfield  {author} {\bibinfo {author} {\bibfnamefont {Z.}~\bibnamefont
  {Fu}}, \bibinfo {author} {\bibfnamefont {C.}~\bibnamefont {Guo}}, \bibinfo
  {author} {\bibfnamefont {B.}~\bibnamefont {Lin}}, \ and\ \bibinfo {author}
  {\bibfnamefont {G.}~\bibnamefont {Liao}},\ }\href@noop {} {\bibfield
  {journal} {\bibinfo  {journal} {Chin. Phys. Lett.}\ }\textbf {\bibinfo
  {volume} {15}},\ \bibinfo {pages} {457} (\bibinfo {year} {1998})}\BibitemShut
  {NoStop}%
\bibitem [{\citenamefont {Zhang}, \citenamefont {Xue},\ and\ \citenamefont
  {Wang}(2002)}]{Zhang2002}%
  \BibitemOpen
  \bibfield  {author} {\bibinfo {author} {\bibfnamefont {D.}~\bibnamefont
  {Zhang}}, \bibinfo {author} {\bibfnamefont {Z.}~\bibnamefont {Xue}}, \ and\
  \bibinfo {author} {\bibfnamefont {Q.}~\bibnamefont {Wang}},\ }\href@noop {}
  {\bibfield  {journal} {\bibinfo  {journal} {J. Phys. D}\ }\textbf {\bibinfo
  {volume} {35}},\ \bibinfo {pages} {2837} (\bibinfo {year}
  {2002})}\BibitemShut {NoStop}%
\bibitem [{\citenamefont {Zhao}\ \emph {et~al.}(2008)\citenamefont {Zhao},
  \citenamefont {Lian}, \citenamefont {Liu},\ and\ \citenamefont
  {Jiang}}]{Zhao2008}%
  \BibitemOpen
  \bibfield  {author} {\bibinfo {author} {\bibfnamefont {L.}~\bibnamefont
  {Zhao}}, \bibinfo {author} {\bibfnamefont {J.-S.}\ \bibnamefont {Lian}},
  \bibinfo {author} {\bibfnamefont {Y.-H.}\ \bibnamefont {Liu}}, \ and\
  \bibinfo {author} {\bibfnamefont {Q.}~\bibnamefont {Jiang}},\ }\href@noop {}
  {\bibfield  {journal} {\bibinfo  {journal} {Transactions of Nonferrous Metals
  Society of China}\ }\textbf {\bibinfo {volume} {18}},\ \bibinfo {pages} {145}
  (\bibinfo {year} {2008})}\BibitemShut {NoStop}%
\bibitem [{\citenamefont {Zhao}\ \emph {et~al.}(2004)\citenamefont {Zhao},
  \citenamefont {Hu}, \citenamefont {Wang}, \citenamefont {Zhao}, \citenamefont
  {Liang},\ and\ \citenamefont {Wang}}]{Zhao2004}%
  \BibitemOpen
  \bibfield  {author} {\bibinfo {author} {\bibfnamefont {J.}~\bibnamefont
  {Zhao}}, \bibinfo {author} {\bibfnamefont {L.}~\bibnamefont {Hu}}, \bibinfo
  {author} {\bibfnamefont {Z.}~\bibnamefont {Wang}}, \bibinfo {author}
  {\bibfnamefont {Y.}~\bibnamefont {Zhao}}, \bibinfo {author} {\bibfnamefont
  {X.}~\bibnamefont {Liang}}, \ and\ \bibinfo {author} {\bibfnamefont
  {M.}~\bibnamefont {Wang}},\ }\href@noop {} {\bibfield  {journal} {\bibinfo
  {journal} {Appl. Surf. Sci.}\ }\textbf {\bibinfo {volume} {229}},\ \bibinfo
  {pages} {311} (\bibinfo {year} {2004})}\BibitemShut {NoStop}%
\bibitem [{\citenamefont {Fujihara}, \citenamefont {Ogawa},\ and\ \citenamefont
  {Kasai}(2004)}]{Fujihara2004}%
  \BibitemOpen
  \bibfield  {author} {\bibinfo {author} {\bibfnamefont {S.}~\bibnamefont
  {Fujihara}}, \bibinfo {author} {\bibfnamefont {Y.}~\bibnamefont {Ogawa}}, \
  and\ \bibinfo {author} {\bibfnamefont {A.}~\bibnamefont {Kasai}},\
  }\href@noop {} {\bibfield  {journal} {\bibinfo  {journal} {Chem. Mater.}\
  }\textbf {\bibinfo {volume} {16}} (\bibinfo {year} {2004})}\BibitemShut
  {NoStop}%
\bibitem [{\citenamefont {Wang}\ \emph {et~al.}(2003)\citenamefont {Wang},
  \citenamefont {Lau}, \citenamefont {Lee}, \citenamefont {Yu}, \citenamefont
  {Tay}, \citenamefont {Zhang},\ and\ \citenamefont {Hng}}]{Wang2003}%
  \BibitemOpen
  \bibfield  {author} {\bibinfo {author} {\bibfnamefont {Y.}~\bibnamefont
  {Wang}}, \bibinfo {author} {\bibfnamefont {S.}~\bibnamefont {Lau}}, \bibinfo
  {author} {\bibfnamefont {H.}~\bibnamefont {Lee}}, \bibinfo {author}
  {\bibfnamefont {S.}~\bibnamefont {Yu}}, \bibinfo {author} {\bibfnamefont
  {B.}~\bibnamefont {Tay}}, \bibinfo {author} {\bibfnamefont {X.}~\bibnamefont
  {Zhang}}, \ and\ \bibinfo {author} {\bibfnamefont {H.}~\bibnamefont {Hng}},\
  }\href@noop {} {\bibfield  {journal} {\bibinfo  {journal} {J. Appl. Phys.}\
  }\textbf {\bibinfo {volume} {94}},\ \bibinfo {pages} {354} (\bibinfo {year}
  {2003})}\BibitemShut {NoStop}%
\bibitem [{\citenamefont {Chen}\ \emph {et~al.}(2002)\citenamefont {Chen},
  \citenamefont {Liu}, \citenamefont {M}, \citenamefont {Zhao}, \citenamefont
  {Zhi}, \citenamefont {Lu}, \citenamefont {Zhang}, \citenamefont {Shen},\ and\
  \citenamefont {Fan}}]{Chen2002}%
  \BibitemOpen
  \bibfield  {author} {\bibinfo {author} {\bibfnamefont {S.}~\bibnamefont
  {Chen}}, \bibinfo {author} {\bibfnamefont {Y.}~\bibnamefont {Liu}}, \bibinfo
  {author} {\bibfnamefont {J.}~\bibnamefont {M}}, \bibinfo {author}
  {\bibfnamefont {D.}~\bibnamefont {Zhao}}, \bibinfo {author} {\bibfnamefont
  {Z.}~\bibnamefont {Zhi}}, \bibinfo {author} {\bibfnamefont {Y.}~\bibnamefont
  {Lu}}, \bibinfo {author} {\bibfnamefont {J.}~\bibnamefont {Zhang}}, \bibinfo
  {author} {\bibfnamefont {D.}~\bibnamefont {Shen}}, \ and\ \bibinfo {author}
  {\bibfnamefont {X.}~\bibnamefont {Fan}},\ }\href@noop {} {\bibfield
  {journal} {\bibinfo  {journal} {J. Cryst. Growth}\ }\textbf {\bibinfo
  {volume} {240}},\ \bibinfo {pages} {467} (\bibinfo {year}
  {2002})}\BibitemShut {NoStop}%
\bibitem [{\citenamefont {Lee}\ and\ \citenamefont {Tu}(2008)}]{Lee2008}%
  \BibitemOpen
  \bibfield  {author} {\bibinfo {author} {\bibfnamefont {M.}~\bibnamefont
  {Lee}}\ and\ \bibinfo {author} {\bibfnamefont {H.}~\bibnamefont {Tu}},\
  }\href@noop {} {\bibfield  {journal} {\bibinfo  {journal} {Jap. J. Appl.
  Phys.}\ }\textbf {\bibinfo {volume} {47}},\ \bibinfo {pages} {980} (\bibinfo
  {year} {2008})}\BibitemShut {NoStop}%
\bibitem [{\citenamefont {Gupta}, \citenamefont {Shridhar},\ and\ \citenamefont
  {Katiyar}(2002)}]{Gupta2002}%
  \BibitemOpen
  \bibfield  {author} {\bibinfo {author} {\bibfnamefont {R.}~\bibnamefont
  {Gupta}}, \bibinfo {author} {\bibfnamefont {N.}~\bibnamefont {Shridhar}}, \
  and\ \bibinfo {author} {\bibfnamefont {M.}~\bibnamefont {Katiyar}},\
  }\href@noop {} {\bibfield  {journal} {\bibinfo  {journal} {Mater. Sci. Semi.
  Proc.}\ }\textbf {\bibinfo {volume} {5}},\ \bibinfo {pages} {11} (\bibinfo
  {year} {2002})}\BibitemShut {NoStop}%
\bibitem [{\citenamefont {Cullity}(1978)}]{Cullity1978}%
  \BibitemOpen
  \bibfield  {author} {\bibinfo {author} {\bibfnamefont {B.}~\bibnamefont
  {Cullity}},\ }\href@noop {} {\emph {\bibinfo {title} {Elements of X-Ray
  Diffractions}}}\ (\bibinfo  {publisher} {Addison-Wesley},\ \bibinfo {address}
  {Reading, MA},\ \bibinfo {year} {1978})\BibitemShut {NoStop}%
\bibitem [{\citenamefont {Jain}, \citenamefont {Sagar},\ and\ \citenamefont
  {Mehra}(2007)}]{Jain2007}%
  \BibitemOpen
  \bibfield  {author} {\bibinfo {author} {\bibfnamefont {A.}~\bibnamefont
  {Jain}}, \bibinfo {author} {\bibfnamefont {P.}~\bibnamefont {Sagar}}, \ and\
  \bibinfo {author} {\bibfnamefont {R.~M.}\ \bibnamefont {Mehra}},\ }\href@noop
  {} {\bibfield  {journal} {\bibinfo  {journal} {Mater. Sci. Pol.}\ }\textbf
  {\bibinfo {volume} {25}},\ \bibinfo {pages} {233} (\bibinfo {year}
  {2007})}\BibitemShut {NoStop}%
\bibitem [{\citenamefont {Tauc}(1968)}]{Tauc1968}%
  \BibitemOpen
  \bibfield  {author} {\bibinfo {author} {\bibfnamefont {J.}~\bibnamefont
  {Tauc}},\ }\href@noop {} {\bibfield  {journal} {\bibinfo  {journal} {Mater.
  Res. Bulletin}\ }\textbf {\bibinfo {volume} {3}},\ \bibinfo {pages} {37}
  (\bibinfo {year} {1968})}\BibitemShut {NoStop}%
\bibitem [{\citenamefont {van Dijken}\ \emph {et~al.}(2000)\citenamefont {van
  Dijken}, \citenamefont {Meulenkamp}, \citenamefont {Vanmaekelbergh},\ and\
  \citenamefont {Meijerink}}]{Dijken2000}%
  \BibitemOpen
  \bibfield  {author} {\bibinfo {author} {\bibfnamefont {A.}~\bibnamefont {van
  Dijken}}, \bibinfo {author} {\bibfnamefont {E.}~\bibnamefont {Meulenkamp}},
  \bibinfo {author} {\bibfnamefont {D.}~\bibnamefont {Vanmaekelbergh}}, \ and\
  \bibinfo {author} {\bibfnamefont {A.}~\bibnamefont {Meijerink}},\ }\href@noop
  {} {\bibfield  {journal} {\bibinfo  {journal} {J. Lumin.}\ }\textbf {\bibinfo
  {volume} {90}},\ \bibinfo {pages} {123} (\bibinfo {year} {2000})}\BibitemShut
  {NoStop}%
\bibitem [{\citenamefont {Aida}\ \emph {et~al.}(2006)\citenamefont {Aida},
  \citenamefont {Tomasella}, \citenamefont {Cellier}, \citenamefont {Jacquet},
  \citenamefont {Bouhssira}, \citenamefont {Abed},\ and\ \citenamefont
  {Mosbah}}]{Aida2006}%
  \BibitemOpen
  \bibfield  {author} {\bibinfo {author} {\bibfnamefont {M.}~\bibnamefont
  {Aida}}, \bibinfo {author} {\bibfnamefont {E.}~\bibnamefont {Tomasella}},
  \bibinfo {author} {\bibfnamefont {J.}~\bibnamefont {Cellier}}, \bibinfo
  {author} {\bibfnamefont {M.}~\bibnamefont {Jacquet}}, \bibinfo {author}
  {\bibfnamefont {N.}~\bibnamefont {Bouhssira}}, \bibinfo {author}
  {\bibfnamefont {S.}~\bibnamefont {Abed}}, \ and\ \bibinfo {author}
  {\bibfnamefont {A.}~\bibnamefont {Mosbah}},\ }\href@noop {} {\bibfield
  {journal} {\bibinfo  {journal} {Thin Solid Films}\ }\textbf {\bibinfo
  {volume} {515}},\ \bibinfo {pages} {1494} (\bibinfo {year}
  {2006})}\BibitemShut {NoStop}%
\end{thebibliography}
\end{document}